\documentclass{aa}
\usepackage{amsmath}
\usepackage{graphicx}
\usepackage{txfonts}
\usepackage{natbib}
\bibpunct{(}{)}{;}{a}{}{,} % to follow the A&A style
\begin{document}
\title{The Wolf-Rayet features and mass-metallicity relation of long-duration
gamma-ray\ burst host galaxies }

\titlerunning{WR features and $M_{*}-Z$ relation of GRB host galaxies}

\author{
   X. H. Han\inst{1,2,3,4}\thanks{email: hxh@nao.cas.cn},
    F. Hammer\inst{2}\thanks{email: francois.hammer@obspm.fr}, Y. C. Liang\inst{3}, H. Flores\inst{2},
    M. Rodrigues\inst{2}, J. L. Hou\inst{1}, J. Y. Wei\inst{3}
    }

\authorrunning{X. H. Han, et al.}

\institute{
Key Laboratory for Research in Galaxies and Cosmology, Shanghai Astronomical
    Observatory, the Chinese Academy of Sciences, 80 Nandan Road, Shanghai,
    200030, PR China
    \and
GEPI, Observatoire de Paris-Meudon, Meudon 92195, France
    \and
National Astronomical Observatories, Chinese Academy of Sciences, Beijing 100012, PR China
    \and
Graduate School of the Chinese Academy of Sciences, Beijing 100049, PR China
             }

\date{Received 12 May 2009 / Accepted 16 January 2010}
%%%%%%%%%%%%%%%%%%%%%%%%%%%%%%%%%%%%%%%%%
\abstract
  % context heading (optional)
  %{} leave it empty if necessary
{}
  % aims heading (mandatory)
{We gather optical spectra of 8 long-duration GRB host galaxies selected from the archival
data of VLT/FORS2. We investigate whether or not
Wolf-Rayet (WR) stars can be detected in these GRB host galaxies. We also
estimate the physical properties of GRB host galaxies, such as metallicity.
   }
  % methods heading (mandatory)
{We identify the WR features in these spectra by
fitting the WR bumps and WR emission lines in blue and red
bumps.
We identify the subtypes of the WR stars,
 estimate the numbers of stars in each subtype, and
calculate the WR/O star ratios.
The (O/H) abundances of
GRB hosts are inferred from both the electron temperature ($T_{e}$) and the
metallicity-sensitive strong-line ratio ($R_{23}$), for which we break the $R_{23}$
degeneracy. We compare the environments of long-duration GRB host
galaxies with those of other galaxies in terms of their luminosity (stellar mass)-metallicity
relations ($L-Z$, $M_{*}-Z$).
 }
  % results heading (mandatory)
{We detect WR stars in 5 GRB host galaxies with spectra of
relatively high signal-to-noise ratios ($S/N$).
In the comparison of $L-Z$, $M_{*}-Z$ relations, we show that GRB hosts
 have lower metallicities
than other samples of comparable luminosity and stellar mass.
The presence of WR stars and the observed high WR/O star ratio,
together with the low metallicity,
support the $\lq\lq$core-collapsar" model and imply that we are witnessing the first stage
of star formation in the host regions of GRBs.
  }
  % conclusions heading (optional), leave it empty if necessary
{}

\maketitle{}

%%%%%TEXT BEGIN%%%%%%%%%%%%%%%%%%%%%%%%%%%%%%%%%%
\section{Introduction} \label{sec_intro}

Gamma-ray bursts (GRBs), the most energetic events in the Universe, were
discovered accidentally in the 1960s~\citep{Klebesadel73}. Since
then, GRBs have been the targets of intense research.  However, the
mechanism behind these bursts and the identity of their progenitors remain disputed.

The durations and spectral properties of GRBs suggest a classification
of short (of duration $\leqslant$ 2 s and hard spectrum)
and long bursts (of duration $\geqslant$ 2 s and
soft spectrum)
~\citep{Kouveliotou93,Hartmann05}.
Short GRBs are believed to originate from the merger of compact binaries such as
double neutron star binaries (NS-NS) and black hole - neutron star binaries (BH-NS).
For long GRBs, the favored $\lq\lq$core-collapsar"
model starts from the idea of a rapidly rotating, massive star that has
undergone extreme gravitational collapse and formed a central black hole
~\citep{Woosley93, MacFadyen99, Klose04}.
According to the collapsar model, the prompt energetic structure of a long GRB
is the result of energy dissipation by internal, relativistic shocks, which may last
seconds or minutes, at a radius of about 10$^{14}$ cm from the center of the
collapsed star \citep[and references therein]{Hartmann05}.
Moreover, the association between GRBs and supernovae (SNe) indicates that in many
cases the parent SN population of GRBs is formed by peculiar type Ibc SNe.

Wolf-Rayet (WR) stars are naturally considered to be the most favored candidates of long
duration GRB progenitors. In the stellar evolution model of~\citet{Hirschi05},
WR stars, which are massive short-lived stars, satisfy the main criteria for GRB
production of black hole formation, loss of hydrogen-rich envelope, and sufficient angular
momentum to form an accretion disk around the black hole.
According to this model, a lower limit to the
metallicity of subsolar value (typically between $\rm{Z_{SMC}}$ and $\rm{Z_{LMC}}$,
i.e., Z $\sim$ 0.2-0.4 $\rm{Z_{\odot}}$) is also a criterion for GRB production.
However, magnetic-field breaking poses some difficulties
in producing GRBs~\citep{Petrovic05}.
To solve this problem, one possible scenario involves assuming that the star
of lower metallicity
rotates so
rapidly that mixing occurs and the star chemically evolves homogeneously
without a hydrogen envelope.
Moreover, its lower metallicity (typically $\rm{Z\lesssim}$ 0.05 $Z_{\odot}$)
causes low
mass loss and therefore the retention of a high angular momentum.
Those conditions are necessary to produce a GRB
\citep{Woosley06, Yoon05}.
Therefore, it is imperative to confirm the presence of WR stars and
determine more accurately metallicities in the region of GRBs.

The subtype and number of WR stars, and
the relative WR/O star number ratio,
are related to the star-forming activity and the starburst duration in galaxies.
We can achieve a deeper understanding of the evolutionary paths of long-duration
GRB progenitors by detecting the WR populations within GRB host galaxies.
Moreover, WR evolutionary models suggest that metallicity, one of the important
diagnostics of the evolutionary histories of galaxies, affects the properties of WR stars
~\citep{Schaerer98}.
\citet{Crowther06} investigated the effect of metallicity on
the WR/O star ratio. They found that the WR/O star ratio
decreases with metallicity. However,
high WR/O star ratios are found in host galaxies of both GRB 980425 and GRB 020903,
 which have low metallicities
~\citep{Hammer06}.  The WR/O star ratio - metallicity relation for other GRB hosts
is still unknown.

The first detection of a counterpart to a GRB at optical and X-ray was achieved on
28 February 1997 ~\citep{Paradijs97}.
Since the discovery of GRB afterglows, GRB host galaxies
and their redshifts have been identifiable.
Between 1997 and 2007, 588 GRBs were detected. Among them, 325 GRBs had X-ray
afterglows, 220 GRBs had optical afterglows,
and 55 GRBs had radio afterglows \footnote{For the most complete list of GRBs, see the
URL: http://www.mpe.mpg.de/$\sim$jcg/grbgen.html, maintained by J. Greiner.}.

The study of GRB host galaxies is very important to the understanding of the
physical properties of
 GRB regions and the nature of GRB progenitors. Although a large number of  host
galaxies can be identified, most of  them are too faint to be observed even using
the largest telescopes in the world. So far, only a few dozens of host galaxies  have
been spectroscopically observed
(e.g., \textbf{GHostS}).
Moreover, the number of GRB hosts that have been
intensively studied spectroscopically is even less.

Evidence from photometric and spectroscopic observations shows that the
host galaxies of long-duration GRBs are mostly faint, blue, low-mass,
star-forming galaxies with low metallicities
\citep[hereafter Savaglio09]{Sokolov01,Floc'h03,Fynbo03,Courty04,Prochaska04, Christensen04,
Chary02, Gorosabel05,Fynbo06, Wiersema07, Kewley07, Levesque09, Savaglio09}.
In contrast,  the host
galaxies of short-duration GRBs mostly have higher luminosities and higher
metallicities than long-duration GRB hosts \citep{Berger08}.

We study a sample of 8 long-duration GRB hosts with high quality spectroscopic observations.
Firstly, we try to detect the WR features in
these GRB hosts then
study the physical properties of GRB host galaxies, such as
 their metallicities and
luminosity (stellar mass)-metallicity relations ($L-Z$, $M_{*}-Z$).
This paper is organized as follows. The sample selection and flux measurements are
performed in Sect.~\ref{sec_data}.  In Sect.~\ref{sec_WR},  we describe the
identification of the WR features in the
spectra of GRB hosts.  In Sect.~\ref{sec_MZ}, the physical properties of GRB host galaxies are
discussed.  The discussion and conclusions
are presented in Sect.~\ref{sec_Dis_Con}.
Throughout the paper, we adopt the $\Lambda$CDM cosmological model  ($H_{0}$ = 70
km s$^{-1}$ Mpc$^{-1}$, $\Omega_{M}$ = 0.3, and $\Omega_{\Lambda}$ = 0.7), and the
 initial mass function (IMF) proposed by \citet{Salpeter55}. All comparisons performed in this work
 include a normalization to the same $\Lambda$CDM cosmological model and IMF.
All magnitudes in this paper are in the Vega system.

%%%%%%%%%%%%%%%%%%%%%%%%%%%%%%%%%%%%%%%
\section{Data reduction and measurements} \label{sec_data}

%%%%%%%%%%%%%%%%%%%%%%%%%%%%%%%%%%%%%%%%

\subsection{The sample selection and data reduction} \label{subsec_sample}

One of the main goals of this work is to find evidence of WR stars in
GRB hosts by means of optical spectroscopic analysis.
We searched archival data of VLT/FORS2\footnote{http://archive.
eso.org/eso/eso-archive-main.html}
to obtain the
spectra of GRB host galaxies at z $<$ 1, which ensures
that a sufficient number of emission lines are included
in the spectral coverage.
These spectra taken with the 600B, 600RI, 600Z
grisms (R $\sim$ 1300), and
300V grism (R $\sim$ 400) of FORS2 were adopted,
because they represent a good compromise between
relatively high spectral resolution, which allows to resolve
the WR features, and relatively high sensitivity, which provides
spectra of high quality.

   \begin{table}
   \setlength\tabcolsep{2pt}
   \centering
   \caption[]{Basic information of the GRB host galaxies}
   \label{Basic information of GRB host galaxies}
   $$
    \begin{array}{lcccccccc}
   \hline
   \noalign{\smallskip}
   \rm{GRB} & \rm{RA^{a}} & \rm{DEC^{a}} & \rm{z} & \rm{E_{G}}(B-V)^b & \rm{Type} \\
   \hline
\rm{GRB~980703}        & 23:59:07 & \rm{ 08:33:36 }& 0.966 &0.061& \rm{long}  \\
\rm{GRB~990712}        &  \rm{ 22:31:50} & \rm{ -73:24:29 }& 0.433 &0.033 & \rm{  long } \\
\rm{ GRB~020405}       & \rm{13:58:03} & \rm{ -31:22:22} & 0.691 &0.055 & \rm{  long } \\
\rm{ GRB~020903}        &  \rm{ 22:49:25} & \rm{ -20:53:59 }& 0.251 & 0.033 & \rm{ long } \\
\rm{ GRB~030329}       & \rm{ 10:44:50} & \rm{ 21:31:17 }& 0.168 &0.025 & \rm{  long}  \\
\rm{ GRB~031203 }      &  \rm{08:02:28} &\rm{ -39:51:04 } & 0.105 & 1.040 & \rm{ long } \\
\rm{ GRB~060218 }       & \rm{03:21:39 }& \rm{ 16:52:02} & 0.034 & 0.140 & \rm{ long } \\
\rm{ GRB~060505  }     & \rm{ 22:07:01} &\rm{ -27:48:56} & 0.089 &0.021 & \rm{long?} \\
   \hline
   \end{array}
   $$
      \begin{description}
\item [Notes:] (a)~Coordinates at the 2000.0 epoch. (b)~ Galactic extinction \citep{Schlegel98}
   \end{description}
   \label{tab_inf}
   \end{table}

%%%%%%%%%%%%%%%%%%%%%%%%%%%%%%%%%%%%%%%%%%%%%%%%%

Data reduction and extraction were performed using a set of
IRAF\footnote{IRAF is distributed by the National Optical
Astronomical Observatories, and is operated by the Association of Universities
for Research in Astronomy, Inc., under cooperative
agreement with the National Science Foundation.} procedures
developed by our team,
which can simultaneously reconstruct
the spectra and the sky counts of
the objects.
For some GRBs associated with supernovae, the early-time
spectra were carefully checked
to avoid the contamination of supernovae, which
increases the difficulty in identifying WR features and the error
in the emission line flux.
Spectra containing contamination by supernovae were then removed.

   \begin{table*}
   \centering
   \caption[]{Spectroscopic observations of GRB host galaxies from VLT/FORS2}
   \label{Observations}
   $$
    \begin{array}{lcccccc}
   \hline
   \noalign{\smallskip}
\rm{UT\ Date}    &  \rm{Exposure\ time(second)}  & \rm{seeing}^{a}('')  &  \rm{Grism}  &  \rm{Program\ ID}\\
   \hline
   \hline
\rm{GRB980703}  &                &      &          & \\
2004.07.15-17   &  8\times1200            &  0.75    &  \rm{600Z}        &  \rm{073.B-0482(A)}\\
2004.07.16-17   &  8\times1200            & 0.75     &  \rm{600RI}       & \rm{073.B-0482(A)}\\
\rm{GRB990712}  &                &      &           & \\
2005.07.05-06   &  4\times1800           & 0.59    &  \rm{300V}        &  \rm{075.D-0771(A)}\\
\rm{GRB020405}  &                &     &           & \\
2004.07.15-16   &  9\times1200        &  0.70      &  \rm{600RI}       &  \rm{073.B-0482(A)}\\
\rm{GRB020903}  &                &    &           & \\
2004.07.15      &  6\times1200             & 0.55     &  \rm{600B}        &  \rm{073.B-0482(A)}\\
2004.07.16      &  6\times1200             & 0.83  &  \rm{600RI}       & \rm{073.B-0482(A)}\\
\rm{GRB030329}  &                 &     &           & \\
2003.04.10      &  6\times600              &  1.18  &  \rm{300V}        &  \rm{071.D-0355(B)}\\
2003.04.17      &  2\times300              &  0.71 &  \rm{300V}        & \rm{071.D-0355(B)}\\
2003.04.22      &  1\times300              &  0.71 &  \rm{300V}        & \rm{071.D-0355(B)}\\
2003.05.01      &  2\times600              &  0.51 &  \rm{300V}        & \rm{071.D-0355(B)} \\
2003.06.19      &  3\times900              &  0.75    &  \rm{300V}        &  \rm{271.D-5006(A)}\\
\rm{GRB031203}  &                 &     &           & \\
2003.12.20      &  2\times2700       & 0.50          &  \rm{300V}        &  \rm{072.D-0480(A)}\\
2004.09.20      &  2\times1800       & 0.60        &  \rm{300V}        &  \rm{073.D-0255(A)}\\
2005.03.28      &  4\times1200       & 0.69          &  \rm{300V}        &  \rm{073.D-0418(A)}\\
2005.04.09      &  3\times1200       & 1.26           &  \rm{300V}        & \rm{075.D-0771(A)}\\
\rm{GRB060218}  &                &      &           & \\
2006.11.27      &  1\times3100       & 0.97           &  \rm{300V}        &  \rm{078.D-0246(A)}\\
2006.12.19-20   &  3\times3150    & 0.72             &  \rm{300V}        & \rm{078.D-0246(A)}\\
\rm{GRB060505}  &                &      &           & \\
2006.05.23      &  2\times1800+1327+600   & 0.81   &  \rm{300V}        &  \rm{077.D-0661(B)}\\
   \hline
   \end{array}
   $$
  \begin{description}
\item [Notes:] (a)~ Seeing is the average value in the observing night.
\end{description}
   \label{tab_obs}
   \end{table*}
%%%%%%%%%%%%%%%%%%%%%%%%%%%%%%%%%%%%%%%%%%%%%%%%%

Wavelength calibration was
performed using the HeNeAr lamp spectra.
Flux calibration was achieved by observing spectrophotometric
 stars with the same grism and the same slit width.
 Galactic extinction
  was adopted from NED\footnote{http://nedwww.ipac.
 caltech.edu/} and \citet{Schaerer98}.
All spectra were transformed into the rest-frame.
The spectra
taken with different grism, but the same resolution
and the same slit width were merged
to widen the wavelength coverages
(e.g., GRB 980703; GRB 020903).
When merging spectra, the exposure times were applied as weighting.
The sample has 8
objects including 7 long-duration GRB hosts
and 1 possible long-duration GRB host (GRB 060505).
The host of GRB 060505 was
spatially resolved by VLT/FORS2 spectroscopic observations,
therefore the GRB region and
the entire host galaxy were studied, separately
\citep[see][for the imaging]{Thone08}.
The detailed discussion about GRB 060505 is given in Sect.~\ref{sec_Dis_Con}

Table~\ref{tab_inf} lists those 8 GRB hosts in our sample with the name, coordinate
 at 2000 epoch, redshift, and burst type.  The detailed observational information
 (date, exposure time, seeing, grism, and program) for
 the objects is given in Table~\ref{tab_obs}.

 %%%%%%%%%%%%%%%%%%%%%%%%%%%%%%%%%%%%%%%%%%

\subsection{Flux measurements of emission lines} \label{subsec_flux}

To measure accurate emission line fluxes, the
continuum and absorption lines should be
subtracted from the spectrum carefully. To do this,
the STARLIGHT\footnote{http://www.starlight,ufsc.br},
a spectral synthesis code, developed by \citet{Fernandes05}, is used to fit the
observed spectrum. STARLIGHT can model the continuum and stellar
absorption lines using a linear combination of
 $N_{*}$ simple stellar population (SSP)
 templates from the evolutionary population synthesis code
 of \citet{Bruzual03}.
In the fitting, we use a base of 45 SSPs, which includes 15
ages (0.001, 0.003, 0.005, 0.01, 0.025, 0.04, 0.1, 0.28, 0.64,
0.9, 1.4, 2.5, 5, 10, 13 Gyr) at 3 metallicities (0.2, 1, 2.5
$Z_{\odot}$).
The Galactic extinction law of \citet{Cardelli89}
(CCM) was adopted in the STARLIGHT fitting.
An example of spectral fitting is shown in Fig.~\ref{12475fig1}.

Emission line fluxes were measured manually using SPLOT
task in IRAF.
The errors originated mainly from three sources:
the first is the uncertainty in fitting both the
continuum and the stellar absorption line;
 the second is the uncertainty in the flux measurement;
the third is the
 Poisson noises from both sky and objects, which dominate the error
 budgets.
Emission line fluxes corrected for
 Galactic extinction are given in Table~\ref{tab_flux}.

%%%%%%%%%%%%%%%%%%%%%%%%%%%%%%%%%%%%%%%%%%%%%%%%
\begin{figure}
 \setcounter{figure}{0}
   \centering
    \includegraphics[bb=10 20 600 450,width=0.52\textwidth]{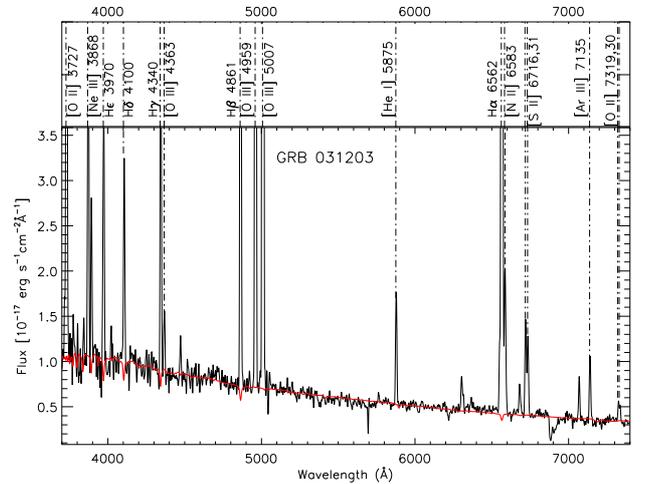}
      \caption{The spectrum of host galaxy of GRB 031203.
      The spectral synthesis fitting is plotted in red line. The strong emission lines are marked.}
    \label{12475fig1}
       \end{figure}

%%%%%%%%%%%%%%%%%%%%%%%%%%%%%%%%%%%%%%%%%%%%%%%%
\begin{table*}
   \setlength\tabcolsep{1pt}
   \tiny
   \centering
   \caption[]{Emission line fluxes of GRB host galaxies  (in units of $10^{-17} \ \rm{erg}~\rm{s}^{-1} ~cm^{-2}$, corrected for Galactic extinction)}
   \label{GRB Host Galaxies emission line fluxes}
   $$
   \begin{array}{ccccccccccccccc}
\hline
\hline
\rm{Ion} & \rm{980703} &  \rm{990712}  &   \rm{020405}  &  \rm{020903}  &  \rm{030329}  &  \rm{031203}    & \rm{060218}  &  \rm{060505^a} & \rm{060505^b}  \\
\hline
\rm{[OII]}\lambda3727  &   24.28\pm0.11   &   23.26\pm0.15   &    11.53\pm0.16   &    11.48\pm0.36   &    19.62\pm0.32  &  943.07\pm4.80  &        224.25\pm1.01  &   20.96\pm0.82   &  50.33\pm1.88 \\
 \rm{H}\delta &    1.66\pm0.22    &    1.95\pm0.09    &   -       &    1.76\pm0.24    &    -       &    204.83\pm4.41  &   20.11\pm0.57   &     - & - \\
\rm{H}\gamma  &    3.1\pm0.19     &    4.9\pm0.10     &    1.9\pm0.09     &   3.94\pm0.29    &    4.41\pm0.10    &   499.73\pm1.10    &   42.02\pm0.67   &   3.79\pm0.30 & - \\
\rm{[OIII]}\lambda4363  &    -     &    0.92\pm0.08  &   -  &    0.55\pm0.07  &    0.71\pm0.05  &    79.10\pm1.25  &    4.77\pm0.30  &    - & - \\
\rm{[FeIII]}\lambda4658 & 0.16\pm0.09 &  -  &  -  &   -  & - & - & - & -  & - \\
\rm{HeII}\lambda4686 & - &  -  &  0.05\pm0.03  &   0.39\pm0.06  & - & 1.20\pm0.92 & 0.81\pm0.31 & - & - \\
\rm{[ArIV]}\lambda4711 & - &  -  &  0.07\pm0.10  &  0.80\pm0.25  & - & 4.90\pm3.11 & 2.12\pm0.84 & - & - \\
\rm{[ArIV]}\lambda4740 &  1.01\pm0.05 &  -  &  -  &   0.41\pm0.10 & - & 3.81\pm2.91 & 2.28\pm0.91 & - & - \\
\rm{H}\beta  &    6.79\pm0.30     &   11.29\pm0.07    &    5.36\pm0.10    &    8.58\pm0.29     &    9.80\pm0.16      &    1135.37\pm5.21      &    91.43\pm0.65    & 8.90\pm0.32 & 21.02\pm0.51 \\
\rm{[OIII]}\lambda4959  &    4.51\pm0.41     &    17.37\pm0.11    &    5.19\pm0.06     &    15.56\pm0.31    &    11.74\pm0.20    &  2404.78\pm5.32  &        108.4\pm0.48    &  8.47\pm0.24 & 12.47\pm0.32  \\
\rm{[OIII]}\lambda5007  &   14.39\pm0.33    &  47.28\pm0.09    &    18.33\pm0.13    &  44.11\pm0.33    &  30.53\pm0.26    &   7238.76\pm4.68  &   291.23\pm0.66   &  20.48\pm0.48 & 27.80\pm2.30 \\
  \rm{H}\alpha  &    -        &   40.34\pm0.32    &    -       &    26.03\pm0.21    &    30.94\pm0.27    &     3636.63\pm4.96  &   261.51\pm0.54   & 31.49\pm0.44 & 114.8\pm3.66  \\
\rm{[NII]}\lambda6583  &    -       &  < 0.41\pm0.20       &    -       &   0.77\pm0.10    &     0.29\pm0.15      &    109.04\pm4.21  &    10.73\pm0.32   & 1.79\pm0.09 & 29.62\pm0.29 \\
\rm{[SII]}\lambda6716  &    -       &   -       &   -       &    2.14\pm0.17    &    3.82\pm0.09    &    129.55\pm4.33  &  15.4\pm0.36    &     3.46\pm0.13 & 26.36\pm0.51 \\
\rm{[SII]}\lambda6731  &    -       &    -       &    -       &   1.28\pm0.09   &   1.89\pm0.08  &   94.63\pm2.89  &  11.56\pm0.41  &     2.80\pm0.20 & 13.98\pm0.19 \\
   \hline
   \end{array}
   $$
\begin{description}
\item [Notes:](a)~GRB site. (b)~The entire host galaxy.
\end{description}
   \label{tab_flux}
   \end{table*}
%%%%%%%%%%%%%%%%%%%%%%%%%%%%%%%%%%%%%%%%%%%%%%%%%

\subsection{Dust extinction} \label{subsec_extinction}

The dust extinction can be estimated from the Balmer-line
ratios (H$\alpha$/H$\beta$, H$\gamma$/H$\beta$,
and H$\alpha$/H$\gamma$).
In our dust extinction calculation, we adopted
Case B recombination with
an electron density of 100 cm$^{-3}$ and a temperature of
10 000 K. The predicted
intrinsic ratio is 2.86 for $I_{0}$(H$\alpha$)/$I_{0}$(H$\beta$),
and 0.466 for $I_{0}$(H$\gamma$)/$I_{0}$(H$\beta$) \citep{Osterbrock89}.
We applied the Galactic extinction law of CCM
for $R_{V}$ = 3.1 where $R_{V}$ is the ratio of $A_{V}$ to $E(B-V)$ \citep{Seaton79}.
The values of dust extinction are listed in Table~\ref{tab_extinction}.
The 2 extinction values for each object derived from the ratios of
H$\alpha$/H$\beta$ and H$\gamma$/H$\beta$ are
 consistent with each other.
The $E(B-V)$ derived from H$\alpha$/H$\beta$ was adopted
in this study when available, because the
higher $S/N$ ratios of these two lines can ensure smaller uncertainties in the results.
The $E(B-V)$ derived from H$\gamma$/H$\beta$
was adopted for GRB 980703 and 020405,
because H$\alpha$ is not available in their spectra.
We then
corrected emission line fluxes for dust extinction.

%%%%%%%%%%%%%%%%%%%%%%%%%%%%%%%%%%%%%%%%%%%%%%%%%

\begin{table}
  \setlength\tabcolsep{1pt}
  \tiny
   \centering
   \caption[]{Dust extinction of GRB host galaxies}
   \label{Dust extinction of GRB host galaxies}
   $$
   \begin{array}{lcccccccccccccc}
   \hline
   \hline
   \noalign{\smallskip}
   \rm{GRB} & \rm{H\alpha}/\rm{H\beta} & \rm{H\gamma}/\rm{H\beta}
   & \rm{A_{V}}(\rm{H\alpha}/\rm{H\beta})  & \rm{A_{V}}(\rm{H\gamma}/\rm{H\beta})
   &\rm{E_{HG}(B-V)}  \\
\hline
980703  &  -  &  0.46\pm0.23  &  -  &  0.14\pm0.53  &  0.05\pm0.17  \\
990712  &  3.57\pm0.41  &  0.43\pm0.11  &  0.55\pm0.03  &  0.50\pm0.15  &  0.18\pm0.01  \\
020405  &  -  &  0.35\pm0.10  &  -  &  1.93\pm0.36  &  0.62\pm0.12  \\
020903  &  3.03\pm0.91  &  0.46\pm0.32  &  0.15\pm0.09  &  0.10\pm0.57  &  0.05\pm0.03  \\
030329  &  3.16\pm0.57  &  0.45\pm0.12  &  0.24\pm0.05  &  0.25\pm0.20  &  0.08\pm0.02  \\
031203  &  3.20\pm1.79  &  0.44\pm0.25  &  0.28\pm0.04  &  0.40\pm0.12  &  0.09\pm0.01  \\
060218  &  2.86\pm1.64 &  0.46\pm0.34  &  0.00\pm0.02  &  0.10\pm0.12  &  0.00\pm0.01  \\
060505^a  &  3.54\pm3.59  &  0.43\pm0.72  &  0.53\pm0.08  &  0.63\pm0.35  &  0.17\pm0.03  \\
060505^b  &  5.46\pm4.51 & - & 1.60\pm0.31 & - & 0.52\pm0.10 \\
   \hline
   \end{array}
   $$
\begin{description}
\item [Notes:](a)~GRB site. (b)~The entire host galaxy.
\end{description}
      \label{tab_extinction}
   \end{table}
%%%%%%%%%%%%%%%%%%%%%%%%%%%%%%%%%%%%%%%%%%%%%%%%%

\section{Wolf-Rayet bump identification} \label{sec_WR}

The main WR features often seen in optical spectra of
galaxies are two characteristic broad emission line clusters.
One is in a blue part of the spectrum at around 4600-4680 $\AA$
\citep[hereafter blue bump,][]{Allen76,Kunth85,Conti91,Schaerer98}.
 The other is in a red region
 around 5650-5800 $\AA$ \citep[hereafter red bump,][]{Kunth86,Dinerstein86}.
The blue bump is actually
a blend of some broad WR lines,
such as $\rm{N\ V}$  4605, 4620  $\AA$, $\rm{N\ III}$  4634, 4640  $\AA$,
$\rm{C\ III}$/$\rm{C\ IV}$ 4650, 4658  $\AA$, and $\rm{He\ II}$ 4686  $\AA$, and some nebular
emission lines superimposed on the bump, such as $\rm{[Fe\ III]}$  4658  $\AA$,
$\rm{He\ II}$  4686 $\AA$, $\rm{He\ I}$ + $\rm{[Ar\ IV]}$  4711  $\AA$, and $\rm{[Ar\ IV]}$  4740
\citep{Guseva00,Izotov98}.
The red bump is usually much weaker than the blue bump.
$\rm{C\ IV}$ 5808  $\AA$ is commonly seen
in the red bump \citep{Kunth86,Dinerstein86}.

The number of WR stars can be estimated from
the luminosity of the blue
and red bumps divided by the luminosity of a single
WR star in a certain subtype
\citep[see, e.g.,][]{Guseva00}. In the following three subsections,
we firstly attempt to identify the WR features in our GRB hosts,
and then to estimate the subtypes and numbers of WR stars, and
finally summarize the
results of these GRB host galaxies.

\subsection{Identifying the WR features}

To be able to identify the WR features reliably,
we select the high quality spectra of GRB hosts.
The criterion for this selection is assumed to be
$S/N$ $>$ 10
in the continuum on both sides of the blue bump to ensure that
the WR features detected are real
and that a final selected sample is
as large as possible.
For other cases, there are very few chances to identify the WR features,
even if they do exist in the spectra.
This final selected sample contains 5 objects:
GRB 980703, GRB 020405, GRB 020903, GRB 031203, and GRB 060218.

We identify the WR features of these 5 GRB hosts based
on the appearance of
the blue bump and WR emission lines by fitting them.
This fitting method is also used by \citet{Brinchmann08}.
The fitting is performed using the SPECFIT routine in IRAF.
We fit the whole blue bump with a single broad
Gaussian profile.
To fit the WR lines $\rm{N\ V}$  4605, 4620  $\AA$, $\rm{N\ III}$  4640  $\AA$,
$\rm{C\ III}$/$\rm{C\ IV}$ 4650, 4658  $\AA$, and $\rm{He\ II}$ 4686  $\AA$, when they
exist,
we adopt multiple Gaussian
profiles to distinguish the broad component, which is mainly
from WR stars, from the narrow component, which is mainly
from nebulae.
Nebular lines,
blended within the bump, are also fitted.
 The $FWHM$s of narrow components of WR lines are
set to be consistent with those of nebular lines.
When carrying out the fit,
the best-fit continuum is chosen very carefully, because the
WR features depend sensitively on the continuum estimates.
We limit the overall wavelength shift of the blue and red features to be
$|\Delta \lambda| \lesssim  3 \AA$ and the width
of the broad component of WR lines to be
$\sim 3000\ \rm{km\,s^{-1}}$ $FWHM$, which are reasonable
values found by \citep{Brinchmann08}.
The fluxes of WR bumps and WR lines are measured from the fitted spectra.
The fits are shown in Fig.~\ref{12475fig2}.
 A significant $\rm{He\ I}$ 5875 $\AA$ emission line, evidence of
young stars such as WR stars, is detected in most of objects.

%%%%%%%%%%%%%%%%%%%%%%%%%%%%%%%%%%%%%%
 \begin{figure*}
   \centering
 \setcounter{figure}{1}
    \includegraphics[width=0.82\textwidth]{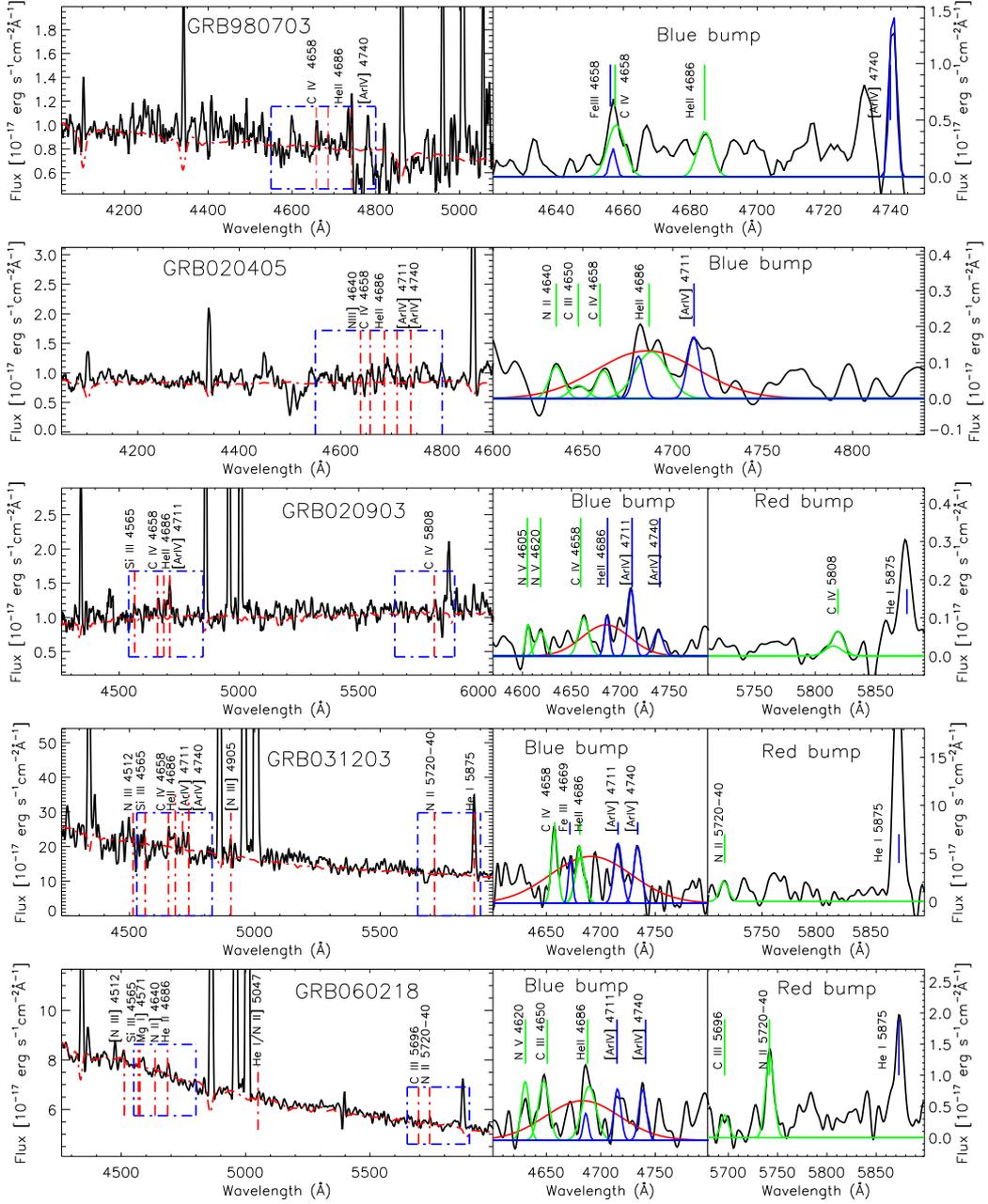}
      \caption{ The spectra of
      GRB hosts are plotted in the left panels. The continuums are marked using
      red lines, which are the results of the Starlight fit using BC03 templates.
       The blue boxes indicate
      the WR blue and red bumps. The WR emission lines are marked
      in red dot-dashed lines. In the right panels, the fits of WR bumps are plotted
      in red profiles. The fits of individual WR line and nebular line are plotted in
      green and blue profiles respectively (for more details, please see the
      online color version.)
}
    \label{12475fig2}
       \end{figure*}
%%%%%%%%%%%%%%%%%%%%%%%%%%%%%%%%%%%%%%%%

\subsection{Identifying the subtypes of WR stars and estimating their numbers}

The WR features in galaxies
originate mainly in two types of WR stars
(WN and WC).
In the WN types, the emission lines from helium and nitrogen ions
are often seen in the spectra.
The emission lines from helium, carbon, and oxygen can be considered as
characteristic features of WC types.
The relative strengths of these emission lines
determine the early (E) and late (L) subtypes of WN stars (WNE, WNL)
\citep[e.g.,][]{Hucht81,Conti83,Vacca92}, and WC stars (WCE, WCL)
\citep[e.g.,][]{Torres86,Vacca92}.
In the blue bump, the broad WR emission lines, such as
$\rm{N\ V}$  4605, 4620  $\AA$, $\rm{N\ III}$  4634, 4640  $\AA$,
$\rm{C\ III}$/$\rm{C\ IV}$ 4650, 4658  $\AA$,
and $\rm{He\ II}$ 4686  $\AA$ are mainly
produced by WNL and WCE stars.
In the red bump, $\rm{C\ IV}$ 5808  $\AA$ is emitted by WCE.
Normally, WNE stars cannot be distinguished from
other WR stars, since they can emit all these lines.
However, their contribution to the bump can be ignored
because of their lower luminosity and
shorter lifetime than WNL stars.
WCL stars can be identified on the basis of their
 $\rm{C\ III}$ 4650, 5696 $\AA$ emission lines.
Another type
of WR star, WO, can be identified using oxygen lines.
Subtypes of WO stars can be classified according to the 
relative strengths of their oxygen lines
\citep{Barlow82,Vacca92}.
In our sample, no object exhibits the features
of WO stars.

The number of WR stars in a certain subtype can be estimated from
the total luminosity of the blue
and/or red bumps divided by the luminosity of a single
WR star \citep{Guseva00}.
To determine accurately the luminosities of the bumps, the contributions from
nebular lines should be subtracted from the bumps.

%%%%%%%%%%%%%%%%%%%%%%%%%%%%%%%%%%%%%%%%%
\begin{table*}
  \small
  \setlength\tabcolsep{2pt}
   \centering
   \caption[]{Properties of WR stars in GRB hosts}
   \label{Properties of WR stars}
$$
\begin{tabular}{c|cc|cc|cc|cc|cc}
\hline
\rm{Ion} &
\multicolumn{2}{c}{980703} &
\multicolumn{2}{c}{020405} &
\multicolumn{2}{c}{020903} &
\multicolumn{2}{c}{031203} &
\multicolumn{2}{c}{060218} \\
\hline
& $F$ & $L$ &   $F$ & $L$ &  $F$ & $L$ &  $F$ & $L$ &  $F$ & $L$ \\
\hline
\rm{blue\ bump}$\ (\lambda4650)^a$ & - & - & 0.18$\pm$0.09 & 654.39$\pm$326.21 & 1.52$\pm$0.31 & 97.22$\pm$42.90 & 10.65$\pm$6.06 & 114.87$\pm$65.37 & 8.08$\pm$3.25 & 6.49$\pm$2.41 \\
\rm{red\ bump}$\ (\lambda5008)$ & - & - & - & - & 0.50$\pm$0.10 & 32.10$\pm$7.30 & - & - & - & - \\
\rm{CIII}$\lambda5696$ & - & - & - & - & - & - & - & - & 3.76$\pm$1.10 & 2.48$\pm$0.76 \\
\hline
WNL &
\multicolumn{2}{c|}{-} &
\multicolumn{2}{c|}{-} &
\multicolumn{2}{c|}{2117$\pm$721} &
\multicolumn{2}{c|}{5743$\pm$3268} &
\multicolumn{2}{c}{324$\pm$120} \\
WCE &
\multicolumn{2}{c|}{-} &
\multicolumn{2}{c|}{-} &
\multicolumn{2}{c|}{1070$\pm$210} &
\multicolumn{2}{c|}{-} &
\multicolumn{2}{c}{-} \\
WCL &
\multicolumn{2}{c|}{-} &
\multicolumn{2}{c|}{-} &
\multicolumn{2}{c|}{-} &
\multicolumn{2}{c|}{-} &
\multicolumn{2}{c}{31$\pm$10} \\
WR/O &
\multicolumn{2}{c|}{-} &
\multicolumn{2}{c|}{0.04$\pm$0.03} &
\multicolumn{2}{c|}{0.21$\pm$0.06} &
\multicolumn{2}{c|}{0.01$\pm$0.01} &
\multicolumn{2}{c}{0.11$\pm$0.04} \\
\hline
\end{tabular}
$$
\begin{description}
\item [Notes:]~WR emission line fluxes ($F$, corrected for Galactic extinction) are in units of $10^{-17} \ \rm{erg}~\rm{s}^{-1} ~cm^{-2}$, Luminosities ($L$) are in units of $10^{38} \ \rm{erg}~\rm{s}^{-1}$.
(a) ~Fluxes and luminosities of nebular lines were subtracted.
\end{description}
\label{tab_wr_flux}
\end{table*}

%%%%%%%%%%%%%%%%%%%%%%%%%%%%%%%%%%%%%%%%%%%%%%%%%

The number of
WCE stars can be derived directly from the luminosity of the red bump,
since that luminosity
 is only produced by WCE stars.
WCE stars are usually
represented by WC4 stars.
The average $\rm{C\ IV}$ 5808  $\AA$ line luminosity of LMC WC stars
 is 3.2 $\times$ $10^{36}$ erg $\rm{s}^{-1}$ \citep{Smith90, Crowther06}.

 To determine
the number of WNL stars, we consider the luminosity of
the blue bump from which we
subtract the contribution of WCE stars.
The contribution of WCE
in the blue bump can be estimated from
the ratio of the luminosity of WC4 stars in the red bump
to that in the blue bump,
%%%%%%%%%%%%%%%%%%%%%%%%%%%%%%%%%%%%%%
\begin{equation} \label{eq_k}
k=\frac{L_{WC4}(\lambda4658)}{L_{WC4}(\lambda5808)}.
\end{equation}
%%%%%%%%%%%%%%%%%%%%%%%%%%%%%%%%%%%%%%%%%%%%%%%%%
Several values of coefficient $k$ were given by previous works.
We adopted the value of $k$ = 1.71 $\pm$ 0.53 given by
\citet{Schaerer98}.
WNL stars are usually
represented by WN7 stars.
The luminosity of a single WN7 in the blue bump,
which depends on metallicity, is 2.0 $\times$ $10^{36}$ erg $\rm{s}^{-1}$
for $Z < Z_{\odot}$, or 2.6 $\times$ $10^{36}$ erg $\rm{s}^{-1}$ for
$Z > Z_{\odot}$ \citep{Schaerer98}.
The first is adopted in this work,
since our samples have lower metallicity than solar (see Sect. 4).

The number of WCL stars can be estimated from the luminosity of
$\rm{C\ III}$ 5696 $\AA$.
For a WCL
single star, the value of luminosity of $\rm{C\ III}$ 5696 $\AA$ adopted in this work is
8.1$\times$ $10^{36}$ erg $\rm{s}^{-1}$
\citep{Schaerer98,Guseva00}.
However, the luminosity of this weak line in a single star is
still not well known.

The ratio of relative WR/O star
number can be estimated by applying two methods to optical spectra.
The method developed by \citet{Arnault89}
uses the flux in the entire WR blue bump, while another
\citep{Vacca92} uses only
the flux of the broad $\rm{He\ II}$ 4686 $\AA$ line.
We use the first method in this work
because it is more suitable for low resolution spectra, for which
the broad $\rm{He\ II}$ 4686 $\AA$ line cannot be separated clearly from the bump.
The formula is
%%%%%%%%%%%%%%%%%%%%%%%%%%%%%%%%%%%%%%%%
\begin{equation} \label{eq_WR/O}
\log{[\frac{WR}{(WR + O)}]}=(-0.11\pm0.02)+(0.85\pm0.02)\log{(\frac{L_{blue\ bump}}{L_{H\beta}})}.
\end{equation}
%%%%%%%%%%%%%%%%%%%%%%%%%%%%%%%%%%%%%%%%
%
For each GRB host, both the fluxes and luminosities of the WR
bumps and other weak WR features are given in Table~\ref{tab_wr_flux}.
Fluxes given here are not corrected for interstellar extinction,
while luminosities are absolute ones, which are transformed from
these fluxes after being corrected for extinction.    

Several sources contribute to the uncertainties
in estimating the numbers of WR stars.
The luminosity of a single WR star adopted is based on 
the average of LMC and Milky Way WR stars, which could be higher 
than our host galaxy sample, because of the difference in metallicity. 
The difference in metallicity could also affect the value of $k$.
Besides these two major effects,
there are three other sources:
the observational uncertainty, the errors in the
multi-component profile fit, and the contamination by
nebular lines to the blue bump.
Together, they produce the measured an
uncertainty, assumed to be 30 - 60~\%, in
the number of WR stars and the
WR/O star number ratio.

\subsection{The results}

We present the WR features and the number of WR stars in each of
our GRB hosts in detail as follows, and as also given in Table~\ref{tab_wr_flux}.

\textbf{GRB 980703} -  The $S/N$ in continuum on both sides of the blue bump (hereafter $S/N$$^*$)
is about 12. We do not
detect the convincing WR bump in the spectrum. However, two WR lines,
$\rm{C\ IV}$ 4658 $\AA$ and $\rm{He\ II}$ 4686 $\AA$ are identified,
which may be emitted by
WNL and WCE stars \citep{Schaerer98,Brinchmann08}.
The red bump is not included within the spectral
range of the spectrum. Therefore, we cannot verify the WR star subtypes.

\textbf{GRB 020405} -  The $S/N$$^*$ is 10.
A blue bump can be seen clearly in the spectrum.
Some broad WR lines, $\rm{N\ II}$ 4640 $\AA$, $\rm{C\ III}$ 4650 $\AA$,
$\rm{C\ IV}$ 4658 $\AA$, and $\rm{He\ II}$ 4686 $\AA$
are identified in the bump,
which implies the presence of WNL and WCE.
The weak $\rm{C\ III}$ 4650 $\AA$ indicates the presence
of WCL stars.
The WR subtypes and the
number of WR stars cannot be verified since there is no
red part to the spectrum.

\textbf{GRB 020903} - The $S/N$$^*$ is about 16.
A strong blue bump is detected in the spectrum.
Broad lines $\rm{N\ V}$  4605, 4620 $\AA$, $\rm{C\ IV}$ 4658 $\AA$,
and $\rm{C\ IV}$ 5808 $\AA$
indicate the existence of WNL and WCE stars \citep{Brinchmann08}.
The number of WCE stars is estimated by using
the luminosity of the red bump,
which can be interpreted as being produced by  1070 $\pm$ 210
WCE stars.
Adopting the value of $k$\ = 1.71 $\pm$ 0.53, the
contribution of these WCE stars in the blue bump is
5.49 $\pm$ 1.81 $\times$ $10^{39}\ \rm{erg}\ \rm{s}^{-1}$.
After subtracting the contribution of WCE stars, the
luminosity of WNL stars in the blue bump is
4.23 $\pm$ 1.46 $\times$ $10^{39}\ \rm{erg}\ \rm{s}^{-1}$,
which corresponds to 2117 $\pm$ 721 WNL stars.

\textbf{GRB 031203} - The $S/N$$^*$ is about 20.
 The strong signature of the blue bump is shown in
the spectrum.  The broad WR lines $\rm{C\ IV}$ 4658 $\AA$, $\rm{He\ II}$ 4686 $\AA$
 are detected in the blue bump.
In the red part, we do not detect the characteristic line of WCE,
$\rm{C\ IV}$ 5808, which implies that the number of WCE in this galaxy can
be ignored. Besides that, the other two WR lines, $\rm{N\ III}$  4905
$\AA$, $\rm{N\ II}$ 5720-40 are also detected.
All of these WR features
are indicative of WNL stars \citep{Smith96,Guseva00,Brinchmann08}.
The number of WNL stars is 1585 $\pm$ 540 as
estimated from the luminosity of the blue bump.

\textbf{GRB 060218} -  The $S/N$$^*$ is higher than 40.
Many WR lines are identified,
 including $\rm{N\ III}$  4512 $\AA$, $\rm{N\ V}$  4620 $\AA$,
$\rm{N\ II}$ 4640 $\AA$, $\rm{He\ II}$ 4686 $\AA$,
$\rm{He\ I}$/$\rm{N\ II}$ 5047 $\AA$, and $\rm{N\ II}$ 5720-40 $\AA$.
All these lines exhibit the characteristics of WNL stars
\citep{Massey03,Crowther97,Guseva00,Brinchmann08}.
Two WR lines, $\rm{C\ III}$ 4650 $\AA$ and $\rm{C\ II}$  5696 $\AA$, are also
identified, which are characteristic of WCL stars \citep{Brinchmann08}.
The presence of WCE stars cannot be confirmed, since no
$\rm{C\ IV}$ 5808 line is detected.
 Using the luminosity of the blue bump,
 we estimate the number of
WNL stars to be 324 $\pm$ 120.
The luminosity of
$\rm{C\ III}$ 5696 $\AA$
corresponds to
31 $\pm$ 10 WCL stars.

The relative WR/O star number ratios are estimated using the luminosities of
blue bumps and H$\beta$. For the hosts of GRB 020405, 020903, 031203, and 060218,
 the ratios are 0.04$\pm$0.03, 0.21$\pm$0.06, 0.01$\pm$0.01,
and 0.11$\pm$0.04, respectively.

In summary, obvious WR features are detected in 4 out of
5 spectra of long-duration GRB hosts, except for GRB 980703, which shows
 no convincing WR bump
 but two WR lines.
 Moreover, the numbers of WR stars for 3 GRB
 hosts are in the range from 300 to 6000, which is consistent
with the typical number of WR stars \citep[$\sim 100-10^5$,][]{Vacca92,Guseva00} 
in WR galaxies. High WR/O star number ratios are also
estimated in these 4 GRB hosts
Therefore, the presence of WR stars in these GRB hosts
is verified.
However, because of the small sample studied here,
the conclusion that WR stars exist in all the long-duration
GRB hosts still cannot be confidently made.
The results of this work
in all cases support
the scenario that WR stars are the progenitors of long GRBs,
and are consistent with the previous studies described below.

\citet{Hammer06} first discovered
WR stars in three nearby long-duration
GRB host  galaxies (GRB 980425, GRB 020903, GRB 031203)
using deep spectroscopic observations.
Since then, little has been done to detect WR features in
long-duration GRB hosts.  \citet{Margutti07} claimed that the identification
of WR emission lines in GRB 031203 is uncertain.
However, their Fig. 3 does show the strong WR blue bump.
The $\rm{N\ III}$  4640 $\AA$ and $\rm{C\ IV}$ 4658 $\AA$ blended broad lines
can also be identified in the bump, which is not attempted in
their work. \citet{Wiersema07} did not detect the obvious
$\rm{He\ II}$ 4686 $\AA$ and WR bump in GRB 060218. However, we note
that the UVES spectra that they used were taken at a time close to the burst, which means
that the significant contamination by the associated SN 2006aj
cannot be ignored in the WR
feature identification.
In their work, an upper limit to the ratio WR/(WR + O)
for the galaxy is estimated to be $<$ 0.4,
which is consistent with the result (WR/O = 0.11$\pm$0.04) estimated in this work.

%%%%%%%%%%%%%%%%%%%%%%%%%%%%%%%%%%%%%%%%%%%%%%%%

\section{The physical properties of GRB host galaxies} \label{sec_MZ}

Metallicity is a fundamental parameter for probing the properties of
GRB progenitors and the environment of GRB regions.
Strong emission lines in spectra of GRB host galaxies allow us to estimate
metallicities.
The luminosity (stellar mass)-metallicity ($L-Z$, $M_{*}-Z$) relation of
galaxies is a fundamental relation for indicating their evolutionary status
and star-formation histories. We estimate the  metallicities of these GRB hosts, and then
study them in terms of the $L-Z$ and $M_{*}-Z$ relations.
%%%%%%%%%%%%%%%%%%%%%%%%%%%%%%%%%%%%%%%%%%%%%%%%

 \subsection{AGN contamination} \label{subsec_bpt}

 The traditional Baldwin-Phillips-Terlevich (BPT) diagram
 \citep{Baldwin81} can diagnose the origin of the narrow emission
  lines of emission-line galaxies.
Fig.~\ref{12475fig3} shows the
  locations of GRB host galaxies in the BPT diagram.
  The $\rm{[O\ III]}$ /H$\beta$ vs. [$\rm{N\ II}$]/H$\alpha$ diagnostic
 lines are taken from \citet{Kewley01} and \citet{Kauffmann03}.
The  $\rm{[O\ III]}$ /H$\beta$ vs. $\rm{[S\ II]}$ /H$\alpha$ relation from
 \citet{Kewley01} is shown.
   This illustrates that
 the GRB host galaxies are star-forming galaxies and AGN contamination in
 GRB host galaxies can be ignored.
%%%%%%%%%%%%%%%%%%%%%%%%%%%%%%%%%%%%%%%%%%%%%%%%

\setcounter{figure}{2}
 \begin{figure}
   \centering
    \includegraphics[bb= 50 20 600 350,width=0.51\textwidth]{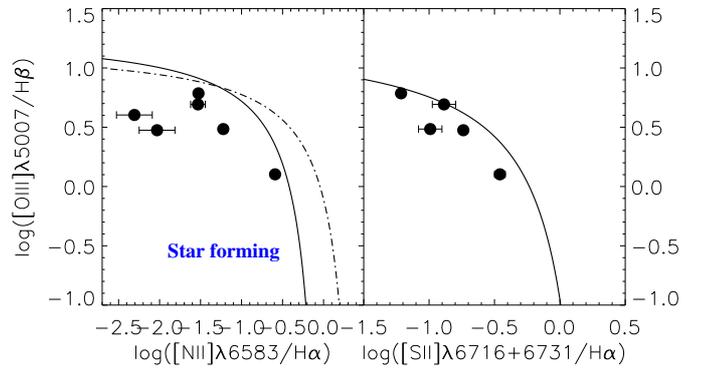}
      \caption{The diagnostic diagrams of GRB host galaxies.
      In the left panel, $\rm{[O\ III]}$ /H$\beta$ vs. [$\rm{N\ II}$]/H$\alpha$ relations is shown.
      The relation from \citet{Kewley01} is drawn in the dot-dashed curve.
      The solid curve is from \citet{Kauffmann03}.
      $\rm{[O\ III]}$ /H$\beta$ vs. $\rm{[S\ II]}$ /H$\alpha$ relation from \citet{Kauffmann03}.
       is shown in the right panel.}
    \label{12475fig3}
       \end{figure}
%%%%%%%%%%%%%%%%%%%%%%%%%%%%%%%%%%%%%%%%%%%%%%%%

\subsection{Metallicity}   \label{subsec_Z}
In terms of oxygen abundance, the metallicities of GRB hosts can be
estimated using the $T_{e}$ method and  $R_{23}$ methods. The $T_{e}$
method  based on
electron temperature is a $\lq\lq$direct" way.
The $\lq\lq$indirect" method is the so-called
$\lq\lq$strong-line" method, such as the $R_{23}$ method based on the empirical
relationship between O/H and $R_{23}$.

It is well known that the $T_{e}$ method is effective
for metal-poor galaxies because the characteristic line
 $\rm{[O\ III]}$4363 is only observable in this
case. Out of 8 galaxies, 5 GRB hosts (GRB 990712, 020903, 030329, 031203,
and 060218) clearly exhibit an
$\rm{[O\ III]}$4363 line, which implies that
they have low metallicities and lie on the lower-metallicity branch of the
$R_{23}$ diagnostic diagram. $\rm{[N\ II]}$/$\rm{[O\ II]}$ also
provides a reliable means of locating
a galaxy on the metallicity branch.
As \citet{Kewley08}
 suggested, the galaxies with
log($\rm{[N\ II]}$6583/$\rm{[O\ II]}$3727) $\leqslant$-1.2
lie on the lower metallicity branch.
All of these 5 objects and the GRB region
of GRB 060505 have values of
log($\rm{[N\ II]}$6583/$\rm{[O\ II]}$3727) $\leqslant$-1.2,
while the entire GRB 060505 host galaxy lies on the upper branch,
because of the value of its log($\rm{[N\ II]}$6583/$\rm{[O\ II]}$3727) $\geqslant$-1.2.
The values of log($\rm{[N\ II]}$6583/$\rm{[O\ II]}$3727) are given in Table~\ref{tab_Z}.
For another 2 GRB hosts, 980703, 020405, we cannot verify
the metallicity branch which on they lie because their [NII]6583
lines are not detected.
Moreover, we use log($\rm{[N\ II]}$6583/H$\alpha$) to verify
the metallicity branches \citep[$\leqslant$-1.3 for lower branch;
$\geqslant$-1.1 for upper branch,][]{Kewley08} for the sample.
These three indicators provide the consistent results for these galaxies.

 %%%%%%%%%%%%%%%%%%%%%%%%%%%%%%%%%%%%%%%%%%
\begin{table*}
   \centering
\caption{Metallicities of GRB host galaxies }
\label{Metallicities of GRB host galaxies}
\setlength\tabcolsep{0.1pt}
\tiny
$$
\begin{array}{lccccccccccc}
   \hline
   \hline
   \noalign{\smallskip}
& 980703 &  990712 &  020405 &  020903 & 030329 & 031203 & 060218 & 060505^g & 060505^h \\
\hline
\rm{N_{e}}[SII]^{a}  & - & 20\sim200 & - & 20\sim 200 & 20\sim 200 & 52.63\pm43.20 & 120.68\pm32.22 & - & -\\
\rm{T_{e}}[OIII]^{b}  & - & 16434.9\pm425.8 & - & 13517.1\pm335.2 & 17911.7\pm325.0 & 12902.2\pm305.2 & 15113.3\pm335.5 & - & -\\
\rm{12+\log{(O/H)} }&\\
(Te)^{c}  & - & 7.812\pm0.15 & - & 8.014\pm0.15 & 7.653\pm0.14 & 8.142\pm0.07 &  7.883\pm0.09 & - & -\\
\rm{12+\log{(O/H)}} &\\
(K99\_l)^{d} & 8.28\pm0.14 & 8.18\pm0.12^f & 8.10\pm0.12 & 8.11\pm0.12^f & 8.06\pm0.10^f & 8.11\pm0.11^f  & 8.16\pm0.14^f & 8.06\pm0.14^f & 7.99\pm0.09 \\
\rm{12+\log{(O/H)}} &\\
 (K99\_u)^{d}  & 8.44\pm0.15 & 8.47\pm0.15 & 8.53\pm0.15 & 8.50\pm0.15 & 8.56\pm0.16 & 8.51\pm0.15 &  8.50\pm0.16 & 8.57\pm0.15 & 8.66\pm0.16^f \\
\rm{\log{([NII]6583}} & \\
/\rm{[OII]3727)} & - & -2.36 & - & -1.47 & -2.12 & -1.23 & -1.45 & -1.36 & -0.52 \\
\rm{A_{V}^{e}} & 0.144\pm0.532 & 0.551\pm0.025 & 1.927\pm0.359 &  0.146\pm0.086 & 0.244\pm0.046 & 0.280\pm0.040 &  0.000\pm0.018 & 0.526\pm0.081 & 1.601\pm0.310 \\
   \hline
   \end{array}
   $$
  \begin{description}
\item [Notes:](a) Electronic density estimated from the
$\rm{[S\ II]}$ flux ratio: $I(6716)/I(6731)$. 
(b) Temperature in K
estimated from the following flux ratios: $\rm{[O\ III]}$
$I(4959+5007)/I(4363)$ 
(c) Metallicity, 12+log(O/H), estimated using the effective
temperature method. 
(d) Metallicity, 12+log(O/H), estimated following
\citet{Kobulnicky99}. K99$\_l$ (their Equation 8) is valid for the
lower branch; K99$\_u$ (their Equation 9) is valid for the upper branch.
(e) Extinction coefficient, $\rm{{A}_V}$ (in magnitude) is derived using
the standard Balmer ratio of H$\alpha$ and H$\beta$. 
For GRB 980703 and 020405, $\rm{{A}_V}$
from the ratio of H$\gamma$ and H$\beta$ is adopted. 
(f) Adopted metallicity in this work. The $R_{23}$ degeneracy are broken
using the $\rm{[N\ II]}$/H$\alpha$ and
 $\rm{[N\ II]}$/$\rm{[O\ II]}$ ratios.
(g) The GRB site
(h) The entire host galaxy
\end{description}
\label{tab_Z}
\end{table*}
%%%%%%%%%%%%%%%%%%%%%%%%%%%%%%%%%%%%%%%%%%%

For the 5 host galaxies that exhibit the $\rm{[O\ III]}$4363 emission line,
we use the $T_{e}$ method to calculate their oxygen abundances.
Electron temperature and electron density are estimated
 using the TEMDEN task of IRAF,
which is based on a 5-level atom program and was described by
\citet{Shaw94}. The flux ratios of
 $\rm{[O\ III]}$ $I(4959+5007)/I(4363)$ are adopted to
calculate electron temperature when available.
Using the flux ratio of
$\rm{[S\ II]}$  $I(6716)/I(6731)$, the electron density can be derived,
which is needed in the electron temperature
and oxygen abundance calculation.

When carrying out the electron density calculation, the
$\rm{[S\ II]}$  $I(6716)/I(6731)$ doublet is not available or too noisy for
some objects in our sample.
Therefore, we tested the sensitivities of both electron
temperature and oxygen abundance to electron density, finding that
neither is very sensitive to electron density.
This allows us to estimate oxygen abundance for a wide range of electron
density (20 - 200 $\rm{cm^{-3}}$) without significant bias (the discrepancy is less than 5 \%).
To calculate $T_{e}$-based oxygen abundances, we use the
methods described by
\citet{Stasinska05}, \citet{Izotov06}, \citet{Yin07}, \citet{Liang06}.

The $R_{23}$ method has been extensively discussed in the literature \citep{Pagel79,McGaugh91,Kobulnicky99,Tremonti04,Yin07,Liang07}.
To estimate their $R_{23}$-based metallicities,
we adopt the formulae of \citet{Kobulnicky99} for both the
upper and lower branches of $R_{23}$-(O/H) solutions.
We use indicators of $\rm{[N\ II]}$6583/$\rm{[O\ II]}$3727
and $\rm{[N\ II]}$6583/H$\alpha$ to break
the $R_{23}$ degeneracy.
The estimated oxygen abundances and errors are listed in Table~\ref{tab_Z}.
The $T_{e}$-based and $R_{23}$-based metallicity estimates
have an acceptable discrepancy of 0.03-0.3dex.

We now compare our metallicity estimates with those in the literature.
There are four objects for which $T_e$-based oxygen abundances
  have been estimated in the literature.
For the host of GRB 020903, \citet{Hammer06} measured a 12+log(O/H)$_{T_e}$ of
7.97, which is very consistent with this work (8.01) within the errorbars;
Savaglio09 gave 8.22, which is 0.21 dex higher.
For the host of GRB 030329, \citet{Levesque09} obtained 7.72 from Keck
spectra, while we obtain 7.65. These two estimates
are extremely consistent within the errorbars.
For the host of GRB 031203,
\citet{Prochaska04} obtained 8.10, \citet{Margutti07} inferred 8.12,
\citet{Levesque09} obtained 7.96,
and Savaglio09 measured 8.02, while we obtain 8.14. All of these estimates
are consistent within the errorbars.

For the host of GRB 060218, \citet{Wiersema07}
measured a 12+log(O/H) value of 7.54,
\citet{Levesque09} obtained 7.62, Savaglio09 measured 7.29,
 and we inferred 7.88, which is closest to the second value.
The discrepancy between these works could be caused by
contamination by the associated supernova.
 Savaglio09 gathered the very early-time spectra observed by \citet{Pian06}
 and \citet{Sollerman06}.
\citet{Wiersema07} acquired their spectrum within 1 month of the burst,
 while our spectrum of the host was taken in Dec. 2006, 10 months after
 the burst, and \citet{Levesque09} took the spectrum even later
 (19 months after the burst).
 In this case, the contamination of the GRB 060218 host galaxy by
 the light of SN2006aj
 was significant and lasted for several months.
This contribution strongly affects measurements of
emission lines; hence, the errors in the metallicity estimates for the early
spectra might be large.
Our result is close to that of \citet{Levesque09}, which
is also derived from the late-time observation.
Anyway, all of them
 show this host has low metallicity.

 We do not compare our $R_{23}$-based estimates with any other work,
 because different calibrations provide very different values, and
 we discussed the discrepancy between our own $T_e$ and $R_{23}$-based
 abundances above.

%%%%%%%%%%%%%%%%%%%%%%%%%%%%%%%%%%%%%%%%%%%%%%%%%
\begin{table*}
   \centering
\caption{Photometry of GRB host galaxies }
$$
   \begin{array}{lccccccccccc}
   \hline
   \hline
   \noalign{\smallskip}
  \rm{GRB} &
  U &
  B &
  V &
  R &
  I &
  J &
  H &
  K &
  \rm{Refs} \\
\hline
980703  &  - & 23.40\pm0.12  &  23.04\pm0.08  &  22.58\pm0.06  &  21.95\pm0.25  &  20.87\pm0.11  &  20.27\pm0.19  &  19.62\pm0.12  &  1, 2, 3 \\
990712  &  23.12\pm0.05  &  23.36\pm0.09  &  22.39\pm0.03  &  21.84\pm0.02 &  21.41\pm0.03  &  20.81\pm0.17  &  20.25\pm0.19  &  20.05\pm0.10  &  3   \\
020903  &  -  &  21.70\pm0.10  &  20.80\pm0.10   &  20.80\pm0.10  &  20.50\pm0.10  &  -    &  -   &  -       &    4, 5    \\
030329  &  22.68\pm0.10  &  23.42\pm0.07  & 22.88\pm0.05  &  22.80\pm0.04  &  -  &  21.52\pm0.04  &  21.17\pm0.24  &  -   &  6  \\
031203  &  22.41\pm0.18  &  22.32\pm0.05  & 20.53\pm0.05 &  20.44\pm0.02  &  19.40\pm0.04  &  18.28\pm0.02  &  17.78\pm0.02  &  16.54\pm0.02  &   7  \\
060218  &  20.45\pm0.15      &  20.46\pm0.07     & 20.19\pm0.04    &  19.86\pm0.03     &  19.47\pm0.06     &  18.99\pm0.16    &  18.52\pm0.22  &  18.69\pm0.34   &  8, 9, 10 \\
060505  &  18.43\pm0.05  &  18.89\pm0.02  &  18.27\pm0.02  &  17.90\pm0.02  &  17.51\pm0.02  &  17.29\pm0.08^a  &  -  &  15.85\pm0.04  &  11  \\
\hline
  & - & - & F555W^a & F702W^a & F814W^a & - & - & - & \\
020405 & - & - & 22.63\pm0.05 & 21.84\pm0.05 & 21.29\pm0.05 & - & - & - & 12 \\
   \hline
   \end{array}
   $$
   \begin{description}
\item [References:] (1)  \citet{Vreeswijk99}; (2) \citet{Sokolov01}; (3) \citet{Christensen04};  (4) \citet{Soderberg04};
(5) \citet{Bersier06};  (6) \citet{Gorosabel05}; (7) \citet{Margutti07}; (8) \citet{Sollerman06};
(9) \citet{Castro08};
(10) \citet{Kocevski07}; (11) \citet{Thone08}; (12) \citet{Wainwright07}
\item [Notes:] The magnitudes are not corrected for Galactic extinction. 
(a) F555W (Central wavelength: 5407 $\AA$; Band width: 1236 $\AA$);
F702W (Central wavelength: 6895 $\AA$; Band width: 1389 $\AA$);
F814W (Central wavelength: 7940 $\AA$; Band width: 1531 $\AA$)
\end{description}
   \label{tab_magnitude}
\end{table*}

%%%%%%%%%%%%%%%%%%%%%%%%%%%%%%%%%%%%%%%%%%%%%%%%%

\subsection{Luminosity $vs.$ metallicity} \label{subsec_L}

\begin{figure}
   \centering
    \includegraphics[bb= 20 20 550 550,width=0.50\textwidth]{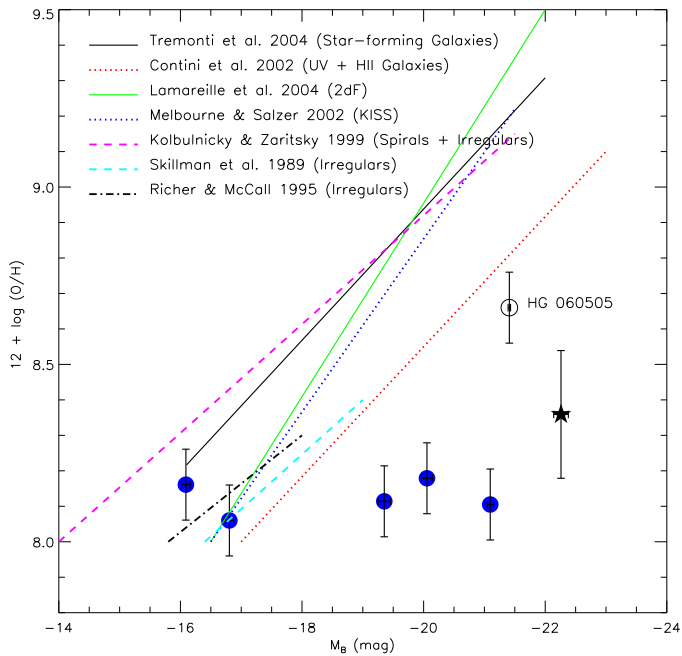}
\caption{ The $L-Z$ relation of long-duration GRB hosts.
The $R_{23}$-based metallicities are adopted.
The blue filled circles represent oxygen abundances which break the degeneracy of $R_{23}$;
      The black star represents the mean value between the lower and upper branch
       of GRB 980703.
      The open circle represents the entire host of GRB 060505.
      Luminosity-metallicity relation for SDSS galaxies and various galaxy samples
      are drawn from the literature (see legend).
      See the online color version for more details.}
    \label{12475fig4}
       \end{figure}

We obtain multi-band photometry from
literature, which is shown in Table~\ref{tab_magnitude}.
All apparent magnitudes are corrected for Galactic foreground extinction
\citep{Schlegel98} and internal extinction.
The $k$-$correction$  is performed for the magnitudes to $z=0$ by using the
$kcorrect$ v4-1-4 program
\footnote{http://cosmo.nyu.edu/~mb144/kcorrect/}
\citep{Oke68,Hogg02,Blanto07}.
The absolute $B$- and $K$-band magnitudes
corrected for Galactic extinction are given
in Table~\ref{tab_stellar_mass}.
We plot our long-duration GRB hosts on the $L-Z$ diagram (Fig.~\ref{12475fig4}),
which is presented
in terms of absolute $B$ magnitude and the $R_{23}$-based metallicities.
The magnitude and metallicity of the entire GRB 060505 host galaxy
are presented in the $L-Z$ diagram.
For GRB 980703,
the mean value between the
lower and upper branch is adopted,
when the $R_{23}$ degeneracy cannot be broken.
The $L-Z$ relations of various
low and high redshift samples from the literature (SDSS star-forming
galaxies from \citet{Tremonti04},
UV-selected galaxies from
\citet{Contini02}, large magnitude-limited sample from \citet{Lamareille04},
emission-line-selected galaxies from \citet{Melbourne02}, irregular
and spiral galaxies from \citet{Kobulnicky_Zaritsky99},
and irregular galaxies from \citet{Skillman89} and \citet{Richer95})
are also shown in Fig.~\ref{12475fig4}.
In comparison with the $L-Z$ relation, the GRB hosts
show an obvious discrepancy from other samples.
GRB hosts have lower metallicity values than other galaxies at given luminosities.
For luminous GRB hosts,
this discrepancy is even larger.
This trend is also shown in \citet{Levesque09}.

%%%%%%%%%%%%%%%%%%%%%%%%%%%%%%%%%%%%%%%%%%%%%%%%%
\begin{table}
\caption{Magnitudes and stellar masses  of GRB host galaxies }
$$
\begin{array}{lccccccccccc}
\hline
   \hline
   \noalign{\smallskip}
  \rm{GRB} &
  \rm{M_B} &
  \rm{M_K} &
  \log{M_* [M_\odot]} \\
\hline
980703  &   -20.85\pm0.12 & -24.41\pm0.12 & 11.14\pm0.39 \\
990712  &   -18.67\pm0.09 & -21.85\pm0.10 &10.14\pm0.34 \\
020405 &    - & - & - \\
020903  &    -18.96\pm0.10 & - & - \\
030329  &    -16.22\pm0.07 & - & - \\
031203  &    -20.58\pm0.05 & -22.27\pm0.02 & 10.13\pm0.26 \\
060218  &   -16.02\pm0.07 & -17.23\pm0.34 & 8.28\pm0.55 \\
060505^a  &   -19.24\pm0.02 & -22.20\pm0.04 & 10.18\pm0.24\\
\hline
\end{array}
$$
\begin{description}
\item [Notes:]Absolute magnitudes $M_B$ and $M_K$
are corrected for galactic foreground extinction \citep{Schlegel98}.
(a) The magnitudes and stellar mass of the entire galaxy of GRB 060505 host
\end{description}
   \label{tab_stellar_mass}
\end{table}
%%%%%%%%%%%%%%%%%%%%%%%%%%%%%%%%%%%%%%%%%%%%%%%%%

%%%%%%%%%%%%%%%%%%%%%%%%%%%%%%%%%%%%%%%%%%%%%%%%%
%%%%%%%%%%%%%%%%%%%%%%%%%%%%%%%%%%%%%%%%%%%%%%%%%

%%%%%%%%%%%%%%%%%%%%%%%%%%%%%%%%%%%%%%%%%%%%%%%%
\subsection{Stellar mass $vs.$ gas phase metallicity} \label{subsec_M}

%%%%%%%%%%%%%%%%%%%%%%%%%%%%%%%%%%%%%%%%%%%%%%%%%
\begin{figure*} \centering \includegraphics[bb= 40 10 560 300,width=14cm]{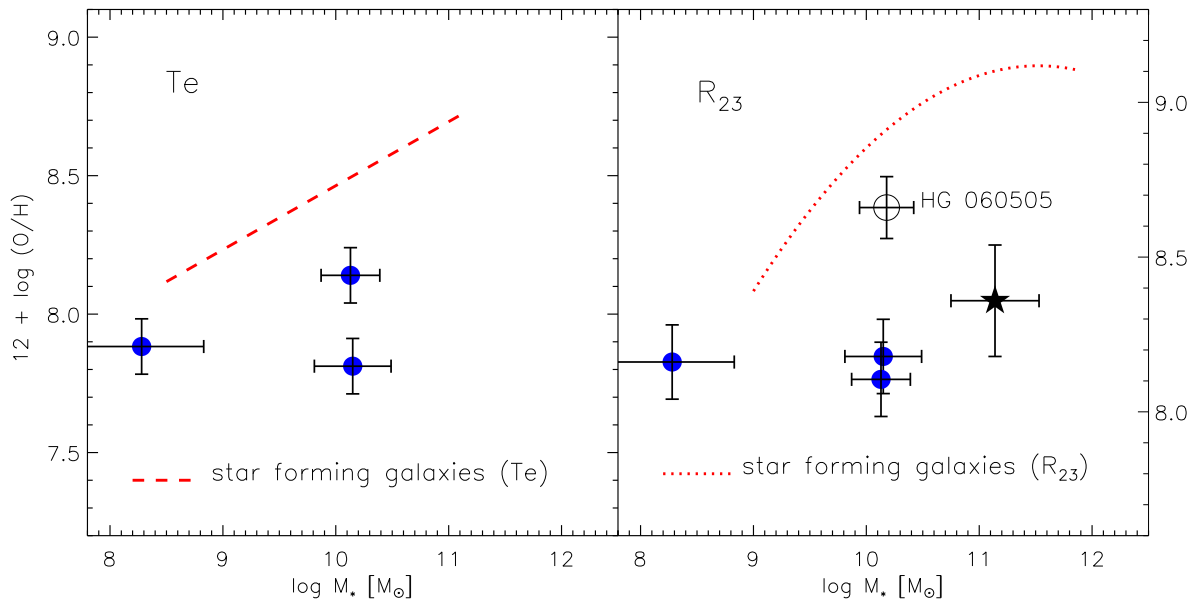}
\caption{ In the left panel,
      the $T_{e}$-based (the filled circles) oxygen abundances of long-duration
      GRB hosts are given.
      The red dashed line refers to the $M_{*}-Z$ relation of local
      star-forming galaxies from SDSS derived from $T_{e}$ method \citep{Liang07}.
      In the right panel, we plot the $R_{23}$-based oxygen
      abundances of long-duration
      GRB hosts derived using formulae from
      \citet{Kobulnicky99} (the blue filled circles - oxygen
      abundances that break the degeneracy;
      the black star - oxygen abundances from mean value between the lower
      and upper branch).
      The open circle represents the entire host of GRB 060505.
      The red dotted line refers to the $M_{*}-Z$ relation of local
      star-forming galaxies from SDSS derived from $R_{23}$ method \citep{Liang07}.
      See the online color version for more details.}
     \label{12475fig5}
      \end{figure*}
%%%%%%%%%%%%%%%%%%%%%%%%%%%%%%%%%%%%%%%%%%%%%%%%%

The stellar mass of galaxy can be estimated from the stellar mass-to-light ratio
and color following \citet{Bell03}.
We use the ($B-V$) color and $M_K$ to estimate the stellar masses of GRB host galaxies.
The formula is
%%%%%%%%%%%%%%%%%%%%%%%%%%%%%%%%%%%%%%%%
\begin{equation} \label{eq_Mstar}
\log({\frac{M_*}{M_{\odot}})=-0.4(M_{K}-3.28)+[a_K+b_K (B-V)+0.15]},
\end{equation}
%%%%%%%%%%%%%%%%%%%%%%%%%%%%%%%%%%%%%%%%
where $M_{K}$ is the $K$-band absolute magnitude and $(B-V)$ is
the rest-frame color. The coefficients
$a_K$ and $b_K$ come from Table 7 of \citet{Bell03}.
The stellar masses are listed in the last column of
Table~\ref{tab_stellar_mass}. We compare our results with previous
work on GRB hosts \citep{Chary02,Castro06,Castro08,Savaglio09}.
  Our results are more consistent with the stellar
masses derived by \citet{Castro08} within the errorbars.
However, our results are systematically higher than theirs
by $\sim$ 0.3 dex.
This discrepancy between the two datasets
could be caused by the use of different $M_{*}/L_{K}$
ratios.
The stellar masses given by Savaglio09 are the lowest of all previous
work.
\citet{Castro08} explain that the reasons for this discrepancy
could be the different $M_{*}/L_{K}$
ratios applied and the underestimated dust extinction in Savaglio09, 
which is also confirmed by
our dust extinction (see Table~\ref{tab_extinction}).

We plot our long-duration GRB hosts on the stellar
mass-metallicity ($M_{*}-Z$) diagram (Fig.~\ref{12475fig5}).
The $T_e$- and $R_{23}$-based metallicities are adopted and
plotted in the left and right panels, respectively.
In the right panel, the open circle and black filled
star represent the hosts of GRB 060505 (not the GRB region) and GRB 980703, respectively.
The $M_{*}-Z$ relation of local normal star-forming galaxies selected from SDSS
 derived by \citet{Liang07} is given in Fig.~\ref{12475fig5}
 (the red dashed line derived from the $T_{e}$ method in the left panel;
the red dotted line derived from the $R_{23}$
 method in the right panel).
 The $T_e$- and $R_{23}$-based metallicities of long-duration GRB host
galaxies are all obviously lower than those of
local star-forming galaxies with comparable stellar masses.

\section{Discussion and conclusions} \label{sec_Dis_Con}

We have attempted to identify the WR stars in GRB host galaxies.
According to the collapsar model, WR stars are considered as the most
favored candidates of the progenitor of long-duration GRBs, therefore
the presence of WR stars in long-duration GRB hosts can directly
validate this model.
We have investigated a sample of
8 GRB host galaxies from archival data from VLT/FORS2.
Out of those 8 GRB hosts, 5 galaxies
 having spectra with $S/N$ $>$ 10 were for our sample
 to detect the WR features.
The presence of WR features was verified
in this entire sample.
Out of those 5 GRB hosts, 4 certainly have WR features detected, which
are GRB 020903,  GRB 031203, GRB 020405, and GRB 060218.
For the remaining one, GRB 980703, the detection is marginal.
The subtypes and numbers of WR stars in those 5 GRB hosts were also derived.
Those results strongly support the collapsar model by illustrating the link
between WR stars and GRBs.

Additional aim was to investigate the physical properties of long-duration GRB
host galaxies, such as metallicity, luminosity, and stellar mass.
Comparing with literature, we
used the most consistent measurements and calculation methods for all
8 objects to ensure systematic coherence and comparability.
We found that the long-duration GRB hosts show obvious
disagreement with the
 $L-Z$ and $M_{*}-Z$ relations derived for low redshift
galaxies in several samples.
The luminous and massive long-duration GRB hosts have
lower metallicities than other
galaxies in these samples.

The WR/O star ratio in WR galaxies is found to decrease as a function of decreasing
metallicity by~\citet{Crowther06}.
The WR/O star ratio and metallicity in GRB host galaxies
do not show a clear trend.
However, compared to
WR galaxies \citep{Guseva00},
the observed WR/O ratios are higher in the GRB hosts studied here, while
metallicities in GRB hosts are
obviously lower than in
the numerous population of low redshift galaxies.
This suggests that the hosted regions of GRBs
are consistent with the first stage
of star formation in a relatively pristine medium.
Furthermore, the ratio of WC/WN stars is also found to decrease with metallicity.
The relation between the ratio of WC/WN stars and metallicity
was derived by \citet{Massey03}.
The hosts of GRB 020903, 060218 deviate from the relation,
in having higher WC/WN ratios
of $\sim$ 0.5 and 0.1 than Local Group galaxies (see their Fig. 11, the upper panel).
The collapsar model suggests that the WC stars are more likely to be
the progenitors of long-duration
GRBs than the WN stars.
Therefore, the observed high WC/WN ratio could also be evidence to
support this model, although the sample is small.

One of our galaxies, the host of GRB 060505, is well resolved in the
ground-based observation.
However,  the type of this burst remains a topic of debate. GRB\,060505
has a burst duration of $\sim$ 4 s, but lacks evidence of an accompanying supernova. It
is classified as a long-duration GRBs that have no
associated supernovae (e.g., GRB 060614) by
\citet{Thone08}. However, \citet{Levesque07}
suggest that the environment of
GRB 060505 is more consistent with the host environments of short-duration GRBs.
We have investigated the metallicity of the GRB site and
the entire host galaxy separately. We have found a relatively low metallicity in the GRB
region and a higher one in the entire host, which are consistent with
\citet{Thone08}.
Unfortunately, the $S/N$ of its spectrum is not high
enough to detect WR star features. However, a very young ($\sim$ 6 Myr)
stellar population in the GRB site is
found by \citet{Thone08} using stellar population modeling.
This low age corresponds to the lifetime of a 32 M$_{\odot}$ star.
Evidence from the properties of the GRB region suggests that the GRB 060505
originated in a long-duration core-collapse progenitor.

Most of the GRB host galaxies in our sample are too distant
to allow spatially resolved analysis
with ground-based spectroscopic observations. Therefore, the only information that we can
derive concerns about the global properties of galaxies.
The properties of the actual explosion
sites are still unclear.
For further study of GRB progenitors and properties of GRB host galaxies,
more high-resolution images and deep spectroscopy are needed. X-shooter at the VLT will be
the ideal instrument to investigate in greater detail the chemical and stellar population properties of GRBs.

%%%%%%%%%%%%%%%%%%%%%%%%%%%%%%%%%%%%%%%%%%%%%%%%%
\begin{acknowledgements}
We thank the anonymous referee for very helpful comments, which improve
well this work.
We are very grateful to P. A. Crowther, X. Y. Chen, H. L. Li, J. Wang, L. P. Xin, and W. K. Zheng for
very helpful discussions.
We acknowledges James Wicker for his careful correcting of English.
This work was supported by the Natural Science Foundation of China
 (NSFC 10933001, 10573028,
 10673002, 10803008, 10821061, 10833005);  the
 National Basic Research Program of China (973 Program, 2007CB815402, 2007CB815404,
 2007CB815406, 2009CB824800); the Young Researcher Grant of National Astronomical
 Observatories, Chinese Academy of Sciences; and the Group Innovation Project (10821302).
 The STARLIGHT project is supported by the Brazilian agencies CNPq,
 CAPES and FAPESP and by the France-Brizil CAPES/Cofecub program.
 \end{acknowledgements}

%%%%%%%%%%%%%%%%%%%%%%%%%%%%%%%%%%%%%%%%%%%%%%%%%

\bibliographystyle{aa} % style aa.bst

\begin{thebibliography}{}
\bibitem[Allen et al.(1976)]{Allen76}
Allen, D. A., Wright, A. E., \& Goss, W. M. 1976, \mnras, 177, 91

\bibitem[Arnault et al.(1989)]{Arnault89}
Arnault, Ph., Kunth, D., \& Schild, H. 1989, \aap, 224, 73

  \bibitem[Baldwin et al.(1981)]{Baldwin81}
  Baldwin, A., Phillips, M. M., \& Terlevich, R. 1981,  \pasp, 93, 817

\bibitem[Barlow \& Hummer(1982)]{Barlow82}
Barlow, M. J., \& Hummer, D. G. 1982, in IAU Symp. 99, Wolf-Rayet Stars:
Observations, Physics, and Evolution, ed. C. W. H. de Loore \& A. J. Willlis
(Dordrecht: Reidel), 387

  \bibitem[Bell et al.(2003)]{Bell03}
  Bell, E. F., McIntosh, D. H., Katz, N., Weinberg, M. D. 2003, \apjs, 149, 289

  \bibitem[Berger(2008)]{Berger08}
  Berger, E.\ 2008, e-prints, arXiv:0805.0306v1

  \bibitem[Bersier et al.(2006)]{Bersier06}
  Bersier, D., Fruchter, A. S., Strolger, L. G., et al. 2006, \apj, 643, 284

  \bibitem[Blanto \& Roweis(2007)]{Blanto07}
  Blanton, M. R., \& Roweis, S. 2007, \aj, 133, 734

\bibitem[Brinchmann et al.(2008)]{Brinchmann08}
Brinchmann, J., Kunth, D., \& Durret, F. 2008, \aap, 485, 657

 \bibitem[Bruzual \& Charlot (2003)]{Bruzual03}
 Bruzual, G., \& Charlot, S. 2003, \mnras, 344, 1000

  \bibitem[Cardelli et al.(1989)]{Cardelli89}
  Cardelli, J. A., Clayton, G. C., \& Mathis, J. S. 1989, \apj, 345, 245

\bibitem[Castro Cer\'on et al.(2006)]{Castro06}
Castro Cer\'on, J. M., Michalowski, M. J., Hjorth, J. et al. 2006, \apj, 653, 85

\bibitem[Castro Cer\'on et al.(2008)]{Castro08}
Castro Cer\'on, J. M., Michalowski, M. J., Hjorth, J. et al. 2008,
e-prints, arXiv: 0803.2235v2

\bibitem[Chary et al.(2002)]{Chary02}
Chary, R., Becklin, E.~E., \& Armus, L.\ 2002, \apj, 566, 229

\bibitem[Christensen et al.(2004)]{Christensen04}
Christensen, L., Hjorth, J., \& Gorosabel, J.\ 2004, \aap, 425, 913

  \bibitem[Cid Fernandes et al.(2005)]{Fernandes05}
  Cid Fernandes, R., Mateus, A., Sodr\'e, L., Stasi\'nska, G.,
  Gomes, J. M. 2005, \mnras, 358, 363

\bibitem[Conti(1991)]{Conti91}
Conti, P. S. 1991, \apj, 377, 115

\bibitem[Conti, Leep \& Perry(1983)]{Conti83}
Conti, P. S., Leep, E. M., \& Perry, D. N. 1983, \apj, 268, 228

\bibitem[Contini et al.(2002)]{Contini02}
Contini, T., Treyer, M. A., Sullivan, M., \& Ellis, R. S. 2002, \mnras, 330, 75

\bibitem[Courty et al.(2004)]{Courty04}
Courty, S., Bj\"ornsson, G., \& Gudmundsson, E. H.\ 2004, \mnras, 354, 581

\bibitem[Crowther \& Hadfield(2006)]{Crowther06}
Crowther, P. A., \& Hadfield, L. J. 2006, \aap, 449, 711

\bibitem[Crowther \& Smith(1997)]{Crowther97}
Crowther, P. A., \& Smith, L. J. 1997, \aap, 320 500

\bibitem[Dinerstein \& Shields(1986)]{Dinerstein86}
Dinerstein, H. L., \& Shields, G. A., 1986, \apj, 311, 45

  \bibitem[Fynbo et al.(2003)]{Fynbo03}
Fynbo, J.~P.~U., Jakobsson, P., M\"oller, P., et al.\ 2003, \aap, 406, L63

  \bibitem[Fynbo et al.(2006)]{Fynbo06}
Fynbo, J.~P.~U., Watson, D., Th\"one, C. C., et al.\ 2006, \nat, 444, 1047

\bibitem[Gorosabel et al.(2005)]{Gorosabel05}
Gorosabel, J., Jel{\'{\i}}nek, M., de Ugarte Postigo, A., Guziy, S., \&
Castro-Tirado, A.~J.,\ 2005, Nuovo Cimento C, 28, 677

\bibitem[Guseva et al.(2000)]{Guseva00}
Guseva, N. G., Izotov, Y. I., \& Thuan, T. X. 2000, \apj, 531, 776

 \bibitem[Hammer et al.(2006)]{Hammer06}
Hammer, F., Flores, H., Schaerer, D., Dessauges-Zavadsky, M.
Le Floc'h, E., \& Puech, M.\ 2006, \aap, 454, 103

 \bibitem[Hartmann(2005)]{Hartmann05}
 Hartmann, D. H. 2005, \nat, 436, 923

  \bibitem[Hirschi et al.(2005)]{Hirschi05}
  Hirschi, R., Meynet, G., \& Maeder, A.\ 2005, \aap, 443, 581

  \bibitem[Hjorth et al.(2005)]{Hjorth05}
  Hjorth, J., Watson, D., Fynbo, J. P. U., et al. 2005, \nat, 437, 859

  \bibitem[Hogg et al.(2002)]{Hogg02}
  Hogg, D. W., Blanton, M., Strateva, I., et al. 2002, \aj, 124, 646

  \bibitem[Izotov et al.(2006)]{Izotov06}
  Izotov, Y. I., Stasi\'nska, G., Meynet, G.,
  Guseva, N. G., Thuan, T. X. 2006, \aap, 448, 955

\bibitem[ Izotov et al.(1998)]{Izotov98}
Izotov, Y. I., \& Thuan, T. X. 1998, \apj, 500, 188


\bibitem[Kauffmann et al.(2003)]{Kauffmann03}
Kauffmann, G. Heckman, T. M., Tremonti, C., et al. 2003, \mnras, 346, 1055

 \bibitem[Kewley et al.(2007)]{Kewley07}
Kewley, L. J., Brown, W. R., Geller, M. J., Kenyon, S. J., \& Kurtz, M. J.\
2007, \aj, 133, 882

\bibitem[Kewley et al.(2001)]{Kewley01}
 Kewley, L. J., Dopita, M. A., Sutherland, R. S., Heisler, C. A., Trevena, J. 2001, \apj, 556, 121

\bibitem[Kewley et al.(2008)]{Kewley08}
 Kewley, L. J., \& Ellison, S. L. 2008, \apj, 681, 1183

\bibitem[Klebesadel, Strong \& Olson(1973)] {Klebesadel73}
Klebesadel, R. W., Strong, I. B., \& Olson, R. A.\ 1973, \apj,  182, 85

 \bibitem[Klose et al.(2004)]{Klose04}
  Klose, S., Greiner, J., Rau, A., et al.\ 2004, \aj, 128, 1942

   \bibitem[Kobulnicky et al.(1999)]{Kobulnicky99}
  Kobulnicky, H. A., Kennicutt, R. C. Jr., \& Pizagno, J. L. 1999, \apj, 514, 544

  \bibitem[Kobulnicky \& Zaritsky(1999)]{Kobulnicky_Zaritsky99}
  Kobulnicky, H. A. \& Zaritsky, D. 1999, \apj, 511, 118

  \bibitem[Kocevski et al.(2007)]{Kocevski07}
  Kocevski, D., Modjaz, M., Bloom, J. S., et al. 2007, \apj, 663, 1180

  \bibitem[Kouveliotou et al.(1993)]{Kouveliotou93}
  Kouveliotou, C., Meegan, C. A., Fishman, G. J., et al. 1993, \apj, 413, 101

 \bibitem[Kunth \& Joubert(1985)]{Kunth85}
Kunth, D., \& Joubert, M. 1985, \aap, 142, 411

\bibitem[Kunth \& Schild(1986)]{Kunth86}
Kunth, D., \& Schild, H. 1986,  \aap, 169, 71

\bibitem[Lamareille et al.(2004)]{Lamareille04}
Lamareille, F., Mouhcine, M., Contini, T., Lewis, L., \& Maddox, S. 2004, \mnras, 350, 396

  \bibitem[Le Floc'h et al.(2003)]{Floc'h03}
  Le Floc'h, E., Duc, P. -A., Mirabel, I. F., et al. 2003, \aap, 400, 499

 \bibitem[Levesque \& Kewley(2007)]{Levesque07}
 Levesque, E. M., \& Kewley, L. J. 2007, \apj, 667, 121

\bibitem[Levesque et al.(2009)]{Levesque09}
Levesque, E. M., Berger, E., Kewley, L. J., \& Bagley, M. M. 2009, e-points, arXiv:0907.4988v3

\bibitem[Liang et al.(2006)]{Liang06}
Liang, Y. C., Hammer, F., Deng, L. C., \& Zhao, G. 2006, PABei, 24, 335

  \bibitem[Liang et al.(2007)]{Liang07}
  Liang, Y. C., Hammer, F., Yin, S. Y., Flores, H., Rodrigues, M., Yang, Y. B. 2007,
  \aap, 473, 411

\bibitem[MacFadyen \& Woosley(1999)]{MacFadyen99}
MacFadyen, A. I., \& Woosley, S. E.\ 1999, \apj, 524, 262

\bibitem[Margutti et al.(2007)]{Margutti07}
Margutti, R., Chincarini, G., Covino, S., et al. 2007, \aap, 474, 815

\bibitem[Massey(2003)]{Massey03}
Massey, P., 2003, ARA\&A, 41, 15

\bibitem[Massey \& Conti(1980)]{Massey80}
Massey, P., \& Conti, P. 1980, \apj, 242, 638

\bibitem[Massey \& Conti(1983)]{Massey83}
Massey, P., \& Conti, P. 1983, \pasp, 95, 440

  \bibitem[McGaugh(1991)]{McGaugh91}
  McGaugh, S. S. 1991, \apj, 380, 140

\bibitem[Melbourne \& Salzer(2002)]{Melbourne02}
Melbourne, J., \& Salzer, J. J. 2002, \aj, 123, 2302

  \bibitem[Oke \& Sandage(1968)]{Oke68}
  Oke, J. B., \& Sandage, A. 1968, \apj, 154, 21

  \bibitem[Osterbrock(1989)]{Osterbrock89}
  Osterbrock, D. E. 1989, Astrophysics of Gaseous Nebulae and Active Galactic Nuclei
  (Mill Valley, California: University Science Books)

  \bibitem[Pagel er al.(1979)]{Pagel79}
  Pagel, B. E., Edmunds, M. G., Blackwell, D. E., Chun, M. S., Smith, G. 1979, \mnras, 189, 95

\bibitem[Petrovic et al.(2005)]{Petrovic05}
Petrovic, J., Langer, N., Yoon, S. -C., Heger, A. 2005, \aap, 435, 247

  \bibitem[Pian et al.(2006)]{Pian06}
  Pian, E., Mazzali, P. A., Masetti, N., et al. 2006, Nature, 442, 1011

   \bibitem[Prochaska et al.(2004)]{Prochaska04}
Prochaska, J.~X., Bloom, J. S., Chen, H. -W., et al.\ 2004, \apj, 611, 200

\bibitem[Richer \& McCall(1995)]{Richer95}
Richer, M. G., \& McCall, M. L., 1995, \apj, 445, 642

\bibitem[Salpeter(1955)]{Salpeter55}
Salpeter, E. E. 1955, \apj, 121, 161

  \bibitem[Savaglio et al.(2009)]{Savaglio09}
  Savaglio, S., Glazebrook, K., \& Le Borgne, D. 2009, \apj, 691, 182

\bibitem[Schaerer \& Vacca(1998)]{Schaerer98}
Schaerer, D., \& Vacca, W. D., 1998, \apj, 497, 618

  \bibitem[Schlegel et al.(1998)]{Schlegel98}
  Schlegel, D. J., Finkbeiner, D. P., \& Davis, M. 1998, \apj, 500, 525

\bibitem[Seaton(1979)]{Seaton79}
Seaton, M. J. 1979, \mnras, 187, 73

\bibitem[Shaw \& Dufour(1994)]{Shaw94}
Shaw, R. A., \& Dufour, R. J. 1994, Astronomical Data Analysis Software and Systems III,
 ed. D. R. Crabtree, R. J. Hanisch, \& J. Barnes, ASP Conf. Ser., 61, 327

\bibitem[Skillman et al.(1989)]{Skillman89}
Skillman, E. D., Kennicutt, R. C., Jr., \& Hodge, P. W. 1989, \apj, 347, 875

\bibitem[Smith et al.(1990)]{Smith90}
Smith, L. F., Shara, M. M., \& Moffat, A. F. J., 1990, \apj, 348, 471

\bibitem[Smith et al.(1996)]{Smith96}
Smith, L. F., Shara, M. M., \& Moffat, A. F. J., 1996, \mnras, 281, 163

\bibitem[Soderberg et al.(2004)]{Soderberg04}
Soderberg, A. M., Kulkarni, S. R., Berger, E., 2004, \apj, 606, 994

  \bibitem[Sokolov et al.(2001)]{Sokolov01}
  Sokolov, V. V., Fatkhullin, T. A., Castro-Tirado, A. J., et al.\ 2001, \aap, 372, 438

  \bibitem[Sollerman et al.(2006)]{Sollerman06}
  Sollerman, J., Jaunsen, A. O., Fynbo, J. P. U., et al. 2006, \aap, 454, 503

  \bibitem[Stasi\'nska(2005)]{Stasinska05}
  Stasi\'nska, G. 2005, \aap, 434, 507

\bibitem[Th\"one et al.(2008)]{Thone08}
Th\"one, C. C., Fynbo, J. P. U., \"ostlin, G., et al. 2008, \apj, 676, 1151

 \bibitem[Tremonti et al.(2004)]{Tremonti04}
  Tremonti, C. A., Heckman, T. M., Kauffmann, G. et al. 2004, \apj, 613, 898

\bibitem[Torres, Conti \& Massey(1986)]{Torres86}
Torres, A. V., Conti, P. S., \& Massey, P. 1986, \apj, 300, 379

 \bibitem[Vacca \& Conti(1992)]{Vacca92}
 Vacca, W. D., \& Conti, P. S. 1992, \apj, 401, 543

 \bibitem[van der Hucht et al.(1981)]{Hucht81}
 van der Hucht, K. A., Conti, P. S., Lundstr\"om, I., \&
 Stenholm, B. 1981, Space Sci. Rev., 28, 227

 \bibitem[van Paradijs et al.(1997)]{Paradijs97}
  van Paradijs, J., Groot, P. J., Galama, T., et al.\ 1997, \nat, 386, 686

\bibitem[Vreeswijk et al.(1999)]{Vreeswijk99}
Vreeswijk, P. M., Galama, T. J., Owens, A., et al. 1999, \apj, 523, 171

\bibitem[Wainwright et al.(2007)]{Wainwright07}
Wainwright, C., Berger, E., \& Penprase, B. E. 2007, \apj, 657, 367

  \bibitem[Wiersema et al.(2007)]{Wiersema07}
Wiersema, K., Savaglio, S., Vreeswijk, P. M, et al.\ 2007, \aap, 464, 529

  \bibitem[Woosley et al.(1993)]{Woosley93}
  Woosley, S. E., Langer, N., \& Weaver, T. A.\ 1993, \apj, 411, 823

  \bibitem[Woosley \& Heger(2006)]{Woosley06}
  Woosley, S. E., \& Heger, A.\ 2006, \apj, 637, 914

  \bibitem[Yin et al.(2007)]{Yin07}
  Yin, S. Y., Liang Y. C., Hammer, F., Brinchmann, J.,   Zhang, B., Deng, L. C., Flores, H.
  2007, \aap, 462, 535

  \bibitem[Yoon \& Langer(2005)]{Yoon05}
  Yoon, S. -C., \& Langer, N. 2005, \aap, 443, 643

\end{thebibliography}
\end{document}